\documentstyle[11pt,gh2001-asp,twoside]{article}
\def\edcomment#1{\iffalse\marginpar{\raggedright\sl#1\/}\else\relax\fi}
\reversemarginpar

\begin{document}
\title{
The Intrinsic Ellipticity of Spiral Disks}
\author{David R. Andersen}
\affil{Max Planck Institute for Astronomy, Heidelberg, Germany}
\author{Matthew A. Bershady}
\affil{Department of Astronomy, UW-Madison, Madison, WI, USA}

\begin{abstract}

We have measured the distribution of intrinsic ellipticities for a
sample of 28 relatively face--on spiral disks. We combine H$\alpha$
velocity fields and $R$ and $I$-band images to determine differences
between kinematic and photometric inclination and position angles,
from which we estimate intrinsic ellipticities of galaxy disks. Our
findings suggest disks have a log-normal distribution of ellipticities
($\overline{\epsilon} =0.06$) and span a range from $\epsilon= 0$
(circular) to $\epsilon=0.2$. We are also able to construct a tight
Tully-Fisher relation for our face-on sample. We use this to assess the
contribution of disk ellipticity on the observed Tully-Fisher scatter.
\end{abstract}

\vspace{-0.25in}
\section{Disk Ellipticity Survey}

Binney (1978) showed triaxial halos could affect disks by inducing
warping and twists.  In particular, the axis ratio of halos lead to
disks which are intrinsically elliptical (Franx \& de Zeeuw 1992; Jog
2000). Hence the ellipticity of disks may plausibly be used to
estimate the axis ratios of dark matter halos.  However, the inability
to disentangle the ellipticity from the phase angle of this distortion
makes such measurements difficult (e.g.  Zaritsky \& Rix 1995;
Schoenmakers 1999).  Andersen et al. (2001) presented a method which
removed this degeneracy and yielded unique solutions for the disk
ellipticity of nearly face-on galaxies by comparing kinematic and
photometric inclination and position angles. This method assumes 
differences between these angles are solely the effect of ellipticity and not
some other distortion. 

Here we present results of a larger study to define the distribution
of disk ellipticities and to establish if ellipticity is related to other
physical quantities, e.g., Tully-Fisher (TF) scatter.  A sample of 39
galaxies were selected from the Principal Galaxy Catalog (Paturel et
al. 1997) which have (1) {\sl
t}-types between 1.5--8.5, (2) axis ratios close to unity, (3)
apparent disk sizes commensurate with the field of view of the
integral field unit, DensePak, on the WIYN 3.5m telescope (Barden,
Sawyer \& Honeycutt 1998), and (4) low galactic absorption. We also required
galaxies in the sample to be unbarred, isolated and have constant
photometric position angles at three scale lengths.

\begin{figure}[t]
\vbox to 2.15in{\rule{0pt}{12in}} \includegraphics{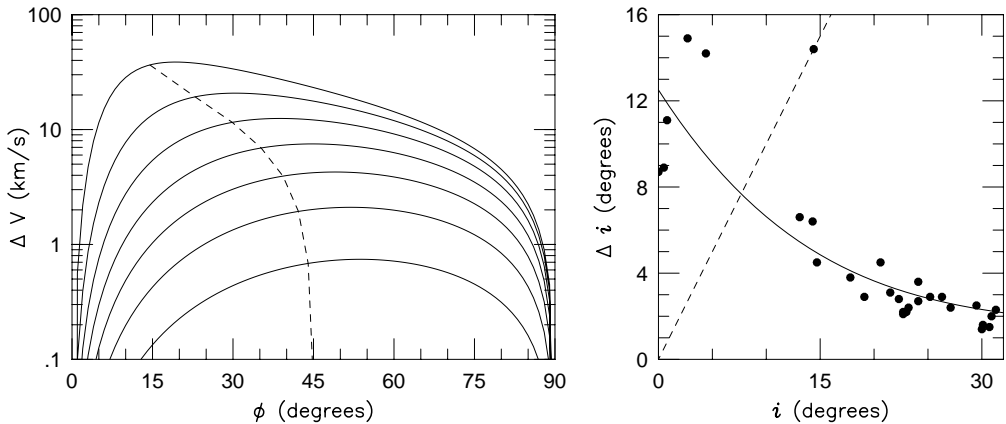}
\caption[]{{\bf Left Panel:} Differences in projected velocity as a
function of $\phi$ (the angle from the major axis in the {\it
observer's} frame) between rotating disks with inclination differences
of 5$^\circ$. These differences assume (1) measurements are made on
the flat part of the rotation curve, (2) $V_{\rm rot}=160 \sin i$ km/s
in the mean, and (3) orbits are circular. The solid curves represent
mean inclinations of $15^\circ, 25^\circ, 35^\circ, 45^\circ,
55^\circ, 65^\circ$ and $75^\circ$. The dashed line represents
$\theta=45^\circ$ for each of these different inclinations, where
$\theta$ is the angle from the major axis in the {\it galaxy}
plane. Classical tilted-ring fits do not utilize data to right of
dashed line, thereby missing over half the signal used to estimate
inclination. {\bf Right Panel:}
The solid curve is our Monte Carlo prediction of inclination errors
for velocity fields ``observed'' with two DensePak pointings and fit
with a single inclined-disk velocity-field model, while points are
errors measured using $\chi^2$ intervals in fits to data. The dashed
line represents $\Delta i/i = 1$. Galaxies with $i>15^\circ$ have
inclination errors $\Delta i< 5^\circ$ which are sufficiently small to
study the TF relation.}
\end{figure}

$R$ and $I$-band images were acquired at the WIYN 3.5m, KPNO 2.1m,
McDonald Observatory 2.7m telescopes.  We used these images to
measure axis ratios and position angles in a way designed to be
unaffected by warps or spiral structure (see Andersen et
al. 2001). H$\alpha$ velocity fields were obtained using DensePak,
feeding the WIYN Bench Spectrograph used with the 316 lines/mm echelle
grating to cover 6500\AA $< \lambda\lambda < 6900$\AA, with an
instrumental FWHM of 0.51 \AA~(22.5 km/s).  Multiple DensePak
pointings allowed us to map H$\alpha$ velocity fields beyond 2.5 disk
scale lengths. We modeled observed velocity fields to derive kinematic
inclinations and position angles --- parameters critical to estimating
disk ellipticity.

\subsection{Velocity-Field Modeling}

Most galaxies in our sample do not show signs of rotation curve
asymmetries, warps, solid body rotation, or spiral structure.
Therefore, we adopted a single, inclined, differentially rotating,
circular disk (``monolithic'') model to fit the DensePak H$\alpha$
velocity fields instead of tilted ring models (e.g., Begeman
1989). There were two major advantages to our approach: (1) A
monolithic velocity-field model uses all data to constrain the fit;
and (2) a monolithic velocity-field model is better able to model
low-inclination disks because tilted ring fits tend to diverge unless
the fit is weighted by $|\cos\theta|$
($\theta$ is an angle measured from a galaxy's major axis
in the galaxy plane) and data with
$|\theta|>\theta_{max} = 45^\circ$ is removed (Begeman 1989). However,
the greatest differences between two velocity-field models with
slightly different inclinations occur at $\theta>45^\circ$, precisely
where tilted-ring fits often do not consider the data (left panel of
Figure 1). Since differences between velocity fields of different
inclination decrease with inclination, it is imperative to use data
at all azimuthal angles to accurately fit velocity-field models at low
inclinations, i.e. $i<30^\circ$ (right panel of Figure 1).  A
hyperbolic tangent rotation curve was sufficient to fit the shape of
rotation curves in our sample with a minimum of free parameters. Our
model had the following free variables: inclination, position angle,
center, central velocity, observed rotation velocity, and hyperbolic
tangent scale-length. The results of the model fits indicate our
approximation that orbits are circular appears
to be acceptable; for $\epsilon_D<0.2$ the model inclination and position
angles derived from
circular versus elliptical orbits would be quite similar. 
We determined kinematic parameters for 36 of 39
galaxies using our fitting procedure.  Of the three galaxies for which
we could not fit velocity-field models, two were at very low
inclinations, while the third had insufficient data.

\subsection{Results}

We derive ellipticities for the 28 of 39 galaxies for 
which accurate measures of the photometric and
kinematic indices exist. 
We find a mean disk ellipticity of
$\overline{\epsilon_D}=0.076$ If we assume the halo potential is
non-rotating and has a constant elliptical distortion, we can estimate
a halo ellipticity of $\overline{\epsilon_\Phi}=0.054$ which is
consistent with previous estimates of halo ellipticity (Rix \&
Zaritsky 1995; Schoenmakers 1999). Our unique solutions for disk
ellipticity also allow us to determine the distribution of
ellipticities for our sample, which we find is well-fit by a
log-normal distribution with a mean and standard deviation on
$\ln(\epsilon_D)$ equal to -2.82$\pm0.73$
($\epsilon_D=0.060^{+0.064}_{-0.031}$, left panel of Figure 2).

\begin{figure}[t]
\vbox to 1.8in{\rule{0pt}{10in}} \includegraphics{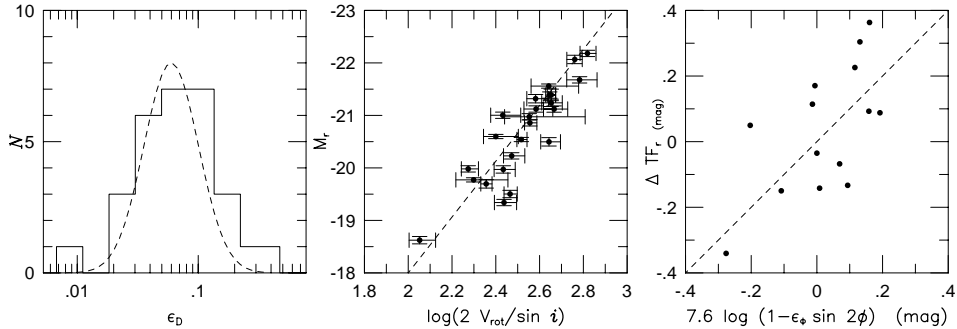}
\caption[]{{\bf Left Panel:} Distribution of disk ellipticities for
our sample of 28 galaxies.  This distribution is well-fit by a
log-normal distribution (dashed line) characterized by a mean ellipticity
$\overline{\epsilon_D}=0.060$.  {\bf Center Panel:} A Tully-Fisher
relation for a sample of galaxies with a mean inclination of
26$^\circ$. The dashed line represents the best fit TF relation to a
subsample of galaxies in the quiet Hubble flow taken from Courteau
(1997) Only 0.44 magnitudes of scatter was exhibited about this
relation.  {\bf Right Panel:} Component of TF scatter due to assuming
circular orbits for an elliptical potential (Franx \& de Zeeuw 1992;
Table 1) versus TF scatter for our sample of nearly face-on galaxies.
}
\end{figure}

\section{Face--On Tully--Fisher Relation}

To demonstrate the precision and reliability of our kinematic
inclinations, we construct a TF relation for 24 galaxies which have
reliable photometry and span a
range of inclinations from 15$^\circ$--35$^\circ$ (a mean of
26$^\circ$).  After applying color corrections, we find our data match
Courteau's (1997) TF relation quite well (central panel of Figure 2).
Courteau's sample galaxies have comparable scale lengths, colors, and
surface brightnesses and distances as ours, but inclinations greater
than 40$^\circ$ (with a mean of 64$^\circ$). Two notable advantages of
using face--on galaxies in TF relation are: Internal absorption
corrections are minimal and the effect of quantities such as
lopsidedness and ellipticity (indices most easily measured in
face-on systems) upon TF scatter can be assessed.

The TF scatter for our sample is quite small (only 0.44 mag) -- quite
similar to the dispersion of Courteau's sample (0.46 mag).  The
distribution of residuals to the TF relation are Gaussian 
except for four galaxies which appear to be outliers. 
There is a correlation between these outliers and kinematic asymmetry:
Of the 5 galaxies in this sample which exhibit strong kinematic
asymmetries in their rotation curves, three are outliers.  If we
exclude the galaxies with strong kinematic asymmetries from our analysis, 
the observed TF scatter is
0.36 mag.

Franx \& de Zeeuw (1992) suggested that disk ellipticity may be a
source of TF scatter.  If we assume a simple, non-rotating model for
the halo potential, we can describe the expected contributions of disk
ellipticity to TF scatter in simple terms. While other astrophysical
sources are expected to contribute a large fraction of the TF error
budget, we do find a statistically significant correlation between
disk ellipticity and TF scatter (right panel of Figure 2).
Constraining the contribution of ellipticity to TF scatter places
limits on other astrophysical sources of TF scatter, including
variations in disk mass-to-light ratios. This work was supported by
NSF grant AST-9970780.

\end{document}